\documentclass[a4paper,UKenglish,cleveref, autoref, thm-restate]{oasics-v2019}

\title{On the Nature of Programming Exercises}
\titlerunning{On the Nature of Programming Exercises}

\author{Alberto Simões}{2Ai, School of Technology, IPCA, Barcelos, Portugal\and\url{https://ambs.zbr.pt}}{asimoes@ipca.pt}{https://orcid.org/0000-0001-6961-2660}{}

\author{Ricardo Queirós}{CRACS, INESC-Porto LA, Portugal \and uniMAD, ESMAD, P.PORTO, Portugal \and \url{http://www.ricardoqueiros.com} }{ricardoqueiros@esmad.ipp.pt}{https://orcid.org/0000-0002-1985-6285}{}

\authorrunning{A. Simões and R. Queirós}
\Copyright{Alberto Simões and Ricardo Queirós}



\funding{This work was partly founded by Portuguese national funds (PIDDAC), through the FCT – Fundação para a Ciência e Tecnologia and FCT/MCTES under the scope of the project UIDB/05549/2020.}

\nolinenumbers 

\EventEditors{John Q. Open and Joan R. Access}
\EventNoEds{2}
\EventLongTitle{42nd Conference on Very Important Topics (ICPEC 2020)}
\EventShortTitle{ICPEC 2020}
\EventAcronym{CVIT}
\EventYear{2016}
\EventDate{December 24--27, 2016}
\EventLocation{Little Whinging, United Kingdom}
\EventLogo{}
\SeriesVolume{42}
\ArticleNo{23}

\begin{document}

\maketitle

\begin{abstract}

There are countless reasons cited in scientific studies to explain the difficulties in programming learning.
The reasons range from the subject's complexity, the ineffective teaching and study methods, 
to psychological aspects such as demotivation. Still, learning programming often boils down to practice on exercise
solving. Hence, it is essential to understand that the nature of a programming exercise is an important
factor for the success and consistent learning.

This paper explores different approaches on the creation of a programming exercise, starting with realizing how
it is currently formalized, presented and evaluated. From there, authors suggest variations that seek to broaden 
the way an exercise is solved and, with this diversity, increase student engagement and learning outcome.
The several types of exercises presented can use gamification techniques fostering student motivation. To
contextualize the student with his peers, we finish presenting metrics that can be obtained by existing automatic assessment tools.
\end{abstract}

\section{Introduction}
\label{sec:introduction}

There is an unanimity regarding the difficulties founded in the teaching-learning process of computer programming. These difficulties are emphasized mainly in introductory teaching, where novice students often lack the knowledge of fundamental programming constructs.
Another explanation is that students, despite being familiar with the constructs, lack the ability of
``problem solving''~\cite{10.5555/1151869.1151901}. Other studies focus on social aspects, since novice students usually
have their introductory programming classes in one of the most difficult periods of their life, that is, at the beginning of a higher education course in computer science, coinciding with a period of transition and instability in their life. There are even authors who consider that the programming courses are not well located in standard computer programming degrees curricula~\cite{Gomes2008AprendizagemDP,Jenkins_onthe}.

In recent years, computer programming training environments appeared with the goal of helping users to learn
programming. The methodology used focus on solving problems from scratch. Nevertheless, initiating the resolution of a
program can be frustrating and demotivating if the student does not know where and how to start.
Based on this fact, some training environments appeared with the support for skeleton programming which facilitates a
top-down design approach, where a partially functional system with complete high-level structures is available. So,
the student needs only to progressively complete or update the code to meet the requirements of the problem. 
Despite its promising results, there are few environments that vary their exercise types in order to motivate 
novice students and keep them focused.

This paper starts by presenting the life cycle of a programming exercise: how it is formalized, how it is presented to the student and how a student's solution is evaluated. Then, the study explores other ways to challenge the student avoiding the ``create from scratch'' assignment.    

The rest of the paper is structured as follows: Section~\ref{sec:pe} explores the current state regarding programming 
exercise formalization and evaluation. Follows Section~\ref{sec:3} where different approaches to construct a programming exercise are analyzed. Finally, the main contributions of the paper and possible paths for future developments are presented.

\section{Programming Exercises}
\label{sec:pe}

The way a programming exercise is formalized and evaluated is crucial for computer programming practice.
In the following subsections we discuss both.

\subsection{Formalization}
\label{sec:formalization}

Until two decades ago, programming assignments were created and presented to students in a \emph{ad hoc} fashion.
The increasing popularity of programming contests worldwide resulted in the creation of several contest management
systems. At the same time Computer Science courses use programming exercises to encourage the practice of programming.
The interoperability between these type of system is becoming a topic of interest in the scientific community. In order 
to address these interoperability issues several formats to represent computer science exercises were 
developed~\cite{DBLP:journals/tlt/QueirosL13}. As notable examples we can found KATTIS~\cite{kattis},
FreeProblemSet\footnote{\url{https://github.com/davideuler/freeproblemset}},
Mooshak Exchange Format~\cite{10.1002/spe.522}, PExIL~\cite{pexil} and YAPeXIL.

The majority of the formats, despite the syntactically differences, adhere to the same logic in terms of structure.
They are based in a XML manifest file referring several types of resources such as problem statements
(e.g. PDF or HTML documents), images, input/output test files, validators (static or dynamic) and solution implementations.
Recently, the YAPExIL, based on PExIL, break these similarities changing the serialization format to JSON 
and supporting different types of programming exercises such as solution improvement, bug fix, gap filling, 
block sorting, and spot the bug\footnote{These types of problems will be discussed in depth in Section~\ref{sec:3}.}. 

\begin{figure*}[!htb]
\centering
\includegraphics[width=0.9\textwidth]{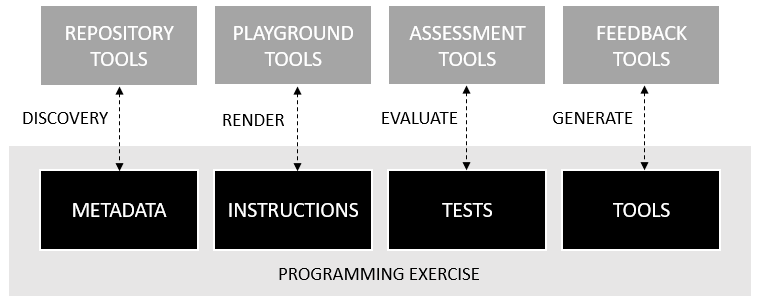}
\caption{Anatomy of a programming exercise.}
\label{fig:architecture}
\end{figure*}

In terms of semantics, all the formats allow the inclusion of:
\begin{itemize}
    \item metadata: data providing information about the exercise. Usually used for discovery actions in repositories; 
    \item instructions: text that is presented to the student (e.g., statement, instructions, skeleton code).
        This data is commonly presented to the student in playground (or training) environments; 
    \item tests: data which is used by the assessment tools to evaluate the student's code.
         The most common data in this category is a set of tests (usually as input/output pairs) and
         a working solution;
    \item tools: tools that the author may use to generate data (e.g. feedback and tests generators, plagiarism tools).
\end{itemize}
In Fig.~\ref{fig:architecture} the four facets and potential tools which will consume the facet data are presented.

\subsection{Evaluation}
\label{sec:evaluation}

The standard way of evaluating a program is to compile it and then execute it with a set of test cases, comprised
of pairs of input/output files. The submitted program is accepted if it compiles without errors and the output of each
run is what is expected.  This is called of dynamic evaluation.

Another approach, is using static evaluation tools that, instead of executing the student's code and injecting input 
data, analyzes the code and a predefined set of metrics is computed. In this context, the presence of a particular
keyword or code block, the style of the code (e.g. variable naming convention), the presence of comments or even the
application of a certain algorithm can be verified. For this type of analysis \emph{linters}, or other static analysis
tools are generally used.

There are several systems that fits on this category such as DOMJudge\footnote{\url{https://www.domjudge.org/}}, 
Mooshak\footnote{\url{https://mooshak.dcc.fc.up.pt/}}, PC$^2$\footnote{\url{https://pc2.ecs.csus.edu/}} and DMOJ\footnote{\url{https://dmoj.ca/}}. 

Most of these systems are contest management systems in the web.
They allow the creation of creating problems, whose solutions can be written in different programming languages,
and have mechanisms to judge automatically the solutions providing (web)interfaces for teams, the jury and the general public.
Some of them (Mooshak and DOMJudge) also provide a REST API to allow their internal functions to be used in other scenarios. All of them are free and open source making them easy to adapt for each one needs.

\section{Types of Programming Exercises}
\label{sec:tpe}\label{sec:3}

There are different approaches to create a programming exercise, depending on what is asked as task for the student, but also in the way the assignment is evaluated and graded. In this section we will first present the different ways a programming exercise can be presented to the student, and what are the main goals of that exercise, and their pros and cons. In some cases examples of application will be discussed. It is followed by an overview of different ways an exercise can be graded, according to the objective. Finally, we will also discuss different ways to give feedback to the students about their performance.

\subsection{Exercise types}
\label{sec:tpe_et}

While a programming exercise can be presented in very different ways, there are some that are traditional and widely used, while some other are rarely applied. These types of exercise not only present a different challenge to the student, but also can be more or less adequate for some specific type of evaluation. And, unfortunately, some types of exercises can take some time for the teacher to prepare.

\subsubsection*{Code from Scratch}
This is the common approach. Easy for the teacher to prepare, as only a statement of a problem is needed. A test suite to evaluate the student’s answer is needed just in the case of using an automatic evaluation tool.
 
For the students, they have a blank sheet, and they will need to code from scratch.  In the student's point of view, this is the worst problem situation. Just like a writer or a painter, they may have the blank page syndrome. There is no indication of where to start. Students can start focusing on the main algorithm to solve the problem, but some students will start with the auxiliary/irrelevant code that is needed, and try to focus later (and probably too late) in the code that the teacher wants to evaluate.

In some situations, like when teaching Object Oriented Programming (for example using Java or C\#), and particularly in the first classes, asking the student to write a static class to be able to write a static main function is counter intuitive, and breaks the Object Orientation logic. 

\subsubsection*{Code Skeleton}
To alleviate the blank page syndrome, and make the student focus on the piece of code being evaluated, the solution is to present a code skeleton, with some blanks to fill in. These can be simple function calls up to complete function or method bodies. Depending on the way this type of problem is presented to the student, the main part of the application might be hidden, and the student will never see the big picture. While this can seem like a bad thing, the fact that the student can focus is a great benefit. The skeleton programs will accelerate the beginning of exercises resolution and facilitate their problem understanding. With the structure included, students can now focus on the core of the problem and abstract their foundations. 

As for the teacher, further work is needed. Teachers will need to write the code skeleton, and present the students with a clear interface, knowing exactly what is available at that point in the program. For complex exercises, teachers might need to write a full solution before being able to understand what pieces of code are to be developed by the student.

\subsubsection*{Fill the Blanks}
Similar to the previous approach, but with smaller blank sections. Students will not need to write full lines or blocks of code, but rather fill in some portions of it. These blanks can be open, allowing the student to write whatever they want, or a predefined list, asking the students to use one of the provided options to fill the blank.

This second approach can be interesting if the students do not have the possibility to run the code, and are presented with very similar options, that will force to really understand what they are performing, without being able to test the code.

In fact, asking students to solve programming tasks without the ability to compile or run their code is relevant, as current compilation times are so fast, that students tend to try all the options/combinations possible for a specific algorithm in order to find the right answer (brute force programming).  

\subsubsection*{Code Baseline}
While in the previous approach the teacher will leave concrete instructions on what code needs to be written, with a code baseline, students will have access to a fully working solution. This working solution might solve the problem for specific values, and students will need to work their code to get a better solution.

This approach is very useful for (but not limited to) teaching how to implement machine learning tools. The teacher can include a solution with a precision baseline, asking the students to accomplish better.

Having a fully working solution, students feel more comfortable as they do not need to write their code from scratch, and feel empowered, as they have a working solution. 
Nevertheless, to start changing the code to get better results, students will need to understand the provided code, and that can be a challenging task, especially if the supplied code is not well documented, or the student is not directed to the code function he needs to change.

As a side benefit from this approach, gamification is implicit. If there is feedback on how well the student’s solution is performing, they will quickly try to beat their friends solutions.

\subsubsection*{Find the Bug}
In this type of exercise the student is asked to merely find the bug (or bugs) for a presented solution. These exercises are used to make students understand an algorithm logic. If the students are in a condition where they cannot compile and test the solution, this is a very interesting approach, as the student is not asked to fix the code.

\subsubsection*{Buggy Code}
Students do not like to rewrite code, trying to make it faster, more elegant, bug free or more generic. The “Find the Bug” type of exercise is a good way to force students to read other’s code, understand it, and change it. They are provided a buggy solution, and need to fix it. 

The types of bugs introduced in the solution can be of different type accordingly with the exercise objectives:
\begin{itemize}
    \item compilation errors: specific syntactic problems are present, like wrong variable types, missing castings, wrong function names, parameter order, etc.
    \item logic errors: the algorithm has serious flaws, and the student needs to detect them. If properly created, these errors can be useful to force the student to understand specific details of an algorithm.
    \item solution errors: the algorithm is mostly working, but have some problems in corner cases. This is similar to the “Baseline” approach, but rather than trying to raise the coverage, precision or accuracy of the solution, the student is asked to make the buggy code work on specific test cases.
\end{itemize}

\subsubsection*{Compiling Errors}
With the spread of intelligent IDEs, students get used to look to the code suggestions, and
very little to the compiler output. An interesting exercise to force students to look and
understand how compilers analyse the code, and how they report syntax errors, is to present the
student with a snippet of code with a syntax error, and the compiler message. This would be
especially interesting if the code snippet is not possible to compile isolated (it uses 
unknown methods) and if the implementation goal is not described. This will force the
student to look carefully to the error message, and to parse the code himself.

\subsubsection*{Code Interpretation}
Just like reading compilation error logs, students lack the ability to understand code. A simple approach to
force students to read and interpret an algorithm is to present the student with a snippet of code, and a set
of options of behaviour. The behaviour can be a description of the algorithm goal, or just information about 
compilation error messages, or faulty behavior. This kind of exercise is interesting if the code is done in a way
the student is not able to copy it and run in a compiler to test it, for example, using non defined functions
described by text.

\subsubsection*{Keyword Use}

This option is an add-on to some specific type of exercises, like the implementation of code from scratch or the
development of a specific function or line of code. In this add-on, the teacher specifies the use of a specific
keyword. As an example, the teacher may require the student to use a map function for a functional style solution
to a specific problem, instead of implementing it as a loop. The main problem on this approach is the automatic
evaluation, because it can not be just a pattern match, as students might use comments to put there the keyword,
or include the keyword in void context, where it does not affect the behaviour of the code. Therefore, the better
approach is to instrument the original function in order to log what was its input, and test there it was 
implemented correctly. The ability to do this kind of instrumentation will depend largely on the used programming
language.

\subsection{Exercise gamification modes}
\label{sec:tpe_egm}

 If different types of exercise test the students knowledge in different situations, gamification increases motivation, challenging their knowledge. Gamification can be introduced just with a ranking on the number, on how many problems were solved by each student, or by assigning (different) points to (different) problems. But this is a rather limited approach to Gamification. Gamification can be used to challenge students to solve a solution in a specific way, and therefore, being not just useful for motivation, but also for learning \cite{DBLP:conf/kes/SwachaQPL19}.

We will discuss how different approaches of gamification can be used to foster learning, and defy students.

\subsubsection*{Slender / Golf}
Instead of just grading a solution accordingly with their result, evaluate the number of characters, instructions, or lines used to solve the problem. In some programming language communities like the Perl Community \cite{GamesDiversionAndPerlCulture} this is seen like a sport, known as Golf or Golfing.

While this challenge is funny, it can be counterproductive. Shorter solutions are usually ugly, difficult to maintain, and explore obscure details of the host programming language. Therefore, while this kind of evaluation can be used with students, it should not be their main goal. 

\subsubsection*{Sprinter}
Efficiency is something students should understand and be able to reach. Teaching them Program Complexity is tedious and non attractive. But if students are challenged to write a fast solution for a problem, they will need to understand the efficiency of different algorithms and data structures in order to score. 

If the solutions are run on a specific hardware (like a server responsible for evaluating the answers), the teacher can prepare a good solution, time it, and define a threshold execution time, forcing students to get their running time below that value.

\subsubsection*{Economic}
Parallel to the Sprinter approach, students are rewarded by the amount of memory they use. Nowadays, given the large amount of memory available on personal computers, students do not have the care to use and reuse memory.

For instance, when teaching the C programming language, it is hard to teach students when they can free memory. This leads to solutions where memory is never freed. Computing the maximum amount of memory used by the solution application during a complete run can be used as a mechanism to motivate students to try to free up memory whenever they do not need it.

\subsubsection*{Sedulous}
Students with learning difficulties can demotivate easily, as they see other students being able to accomplish working and probably fast and economic solutions. Rewarding students that attempt to solve a problem more than a fixed amount of tries can be motivational. Of course that the grading system should be able to understand if those are honest attempts or if the student is just trying to send always one wrong solution just to be rewarded.   

\subsubsection*{Scout}
Provides a bonus reward when the student makes several tests to check his solution, before submitting.
This is not something that can be easily accounted for automatically. A good alternative is to give a bonus
to the student if it passes all the tests with the first submitted solution.

\subsubsection*{Meticulous}
Sometimes there are different ways to accomplish a working solution, and the number of lines, the code efficiency
or amount of used memory is not enough to distinguish the chosen solution approach. With this in mind, teachers
can define a set of specific keywords or function/methods that will give a bonus to the student's solution.

The main problem for this solution is the possibility of cheating. If the student gets aware that a specific keyword
is being checked, he might just write that keyword in a comment, or in a void context, where it is not exactly being
used as it should. A way to circumvent this cheating approach is to hide to the student how the grading system works,
or to define a wrapper to the functions being tested, that evaluate how they are being used in the student's solution.

\subsection{Exercise Statistics}

In the previous section we presented some ways to grade students accordingly with different factors that do not
relate necessarily with the correctness of the solution. Teachers might not want to use all of those grading
approaches at the same time. Nevertheless, computing statistics on some of the presented factors, and showing 
them to the students can work, indirectly, as gamification.

Thus, it is suggested to add solution metrics regarding each problem submitted solutions. Follow some simple
examples:

\begin{itemize}
    \item Average Solution Time: how much time a student takes to prepare a solution, starting from the
    moment the problem description was seen, up to a good solution to be submitted. This will allow students to
    understand how they relate with their mates, and will allow the teacher to understand how his students
    problem solving abilities are.
    \item Wrong Attempts Average: how many attempts (in average) a student performs, before getting the
    solution accepted. If this number is high, students might not have understood the problem correctly,
    or they are trying at force to get a solution, instead of really thinking in a good approach.
    \item Least Memory Used: who is the student having the solution using less memory for each problem.
    \item Shortest Execution Time: who is the student having the fastest solution.
    \item Average Execution Time: what is the average execution time for a specific problem solutions.
\end{itemize}

\section{Conclusions}
Learning programming is a difficult task. Many reasons have been shared among the scientific community. However, it is important not to forget, that learning programming requires constant practice. In programming, this practice boils down to solving exercises, often from scratch. While this could be simple for average and expert students, for novice students this approach can negatively affect his performance in the course and, consequently, increase their demotivation. Therefore, this paper describes several types of exercises in order to cover different learning profiles and enhance new skills. This diversity is seen by the authors as beneficial to not making the challenges tedious for more advanced students and to support novice students to consolidate their skills.

As future work, the authors will try to explore simple ways to facilitate the process of changing an exercise type through standard and non-language dependent techniques.

\label{sec:conclusions}

%
%
\bibliography{onpe}

\end{document}